\documentclass[10pt,letterpaper]{article}
\usepackage[top=0.85in,left=1.6in,footskip=0.75in,marginparwidth=2in]{geometry}

\usepackage[utf8]{inputenc}
\usepackage{ragged2e}
\usepackage{siunitx}
\usepackage{cite}
\usepackage{nameref,hyperref}
\usepackage{amsmath}
\usepackage{booktabs} 
\usepackage{microtype}
\DisableLigatures[f]{encoding = *, family = * }

\raggedright
\usepackage{changepage}
\usepackage[aboveskip=1pt,labelfont=bf,labelsep=period,singlelinecheck=off]{caption}

\makeatletter
\renewcommand{\@biblabel}[1]{\quad#1.}
\makeatother

\usepackage{lastpage,fancyhdr,graphicx}
\usepackage{epstopdf}
\fancyheadoffset[L]{2.25in}
\fancyfootoffset[L]{2.25in}

\usepackage{color}

\definecolor{Gray}{gray}{.25}

\usepackage{graphicx}

\usepackage{sidecap}

\usepackage{wrapfig}
\usepackage[pscoord]{eso-pic}
\usepackage[fulladjust]{marginnote}
\reversemarginpar

\begin{document}
\vspace*{0.35in}

\begin{flushleft}
{\Large
\textbf\newline{pH-Sensitive Ultra-thin Oxide-Liquid Metal System: Understanding the Fundamental Sensing Mechanism}
}
\newline
\\
Atanu Das\textsuperscript{1}

\bigskip
\bf{1} Department of Electronics and Communication Engineering, Manipal Institute of Technology, Manipal Academy of Higher Education, Manipal, India
\\
\bigskip
atanu.das@manipal.edu

\end{flushleft}
\section*{Abstract}
\justifying
The pH response of liquid metal (eutectic GaInSn) in the form of a pendant drop is investigated and the sensitivity of (92.96±13.54) mV $pH^{-1}$ in the pH range from 4 to 10 is obtained. Unlike the fundamental limit of pH sensitivity of 59.1 mV $pH^{-1}$ in an electrolyte-site binding surface, the super-Nernstian pH sensitivity originated from a spontaneous electrochemical reaction associated with an enhanced ionic exchange at the ultra-thin (1-3 nm) $Ga_{2}O_{3}$-electrolyte interface which is purely driven by thermodynamics, rendering to the lowest system energy possible involving gallate ($GaO_{3}^{-}$, $GaO_{3}^{3-}$) and bi-gallate ($HGaO_{3}^{2-}$) ions. A unified Nernst equation is derived by introducing an ion-exchange factor “$x$” ($x=n_{H^{+}}/n_{e^{-}}$) to explain super-Nernstian pH sensitivity and found a direct link between pH sensitivity and Pourbaix ‘pH-Potential’ formulations. It is found that Nernstian sensitivity of 59.1 mV $pH^{-1}$ occurs only for symmetric ion exchange ($x=1$) reaction, whereas asymmetric ion exchanges ($x\neq 1$) could result in sensitivity far beyond the Nernst sensitivity. Our findings have great scientific significance, which could redefine the conventional concept of the ion sensing mechanism in a solid-state electrochemical sensor and push forward the future development of the 2D oxide-based electrochemical sensor.


\section*{Introduction}
 The discovery of graphene from natural graphite in 2004\cite{RN1} resonates with the world's scientific community and opens up research opportunities in 2D materials. Atomically smooth 2D natural superficial oxide on liquid gallium and its alloys was isolated in 2017\cite{RN2} which expands the horizon of 2D material. Gallium and its eutectic alloys $EGa_{0.68}In_{0.21}Sn_{0.10}$  have high electrical and thermal conductivity, excellent fluidity and flexibility, non-toxicity, and bio-compatibility.  An ultra-thin oxide (0.5$\sim$3 nm) formation\cite{RN3} on liquid metal with ambient air or a trace amount of oxygen ($<1 ppm)$ gives the liquid metal an added attribute, which enables its application much more appealing to the scientific community.\cite{RN4, RN5} In recent times, liquid metal research gets substantial momentum with several noteworthy findings,\cite{RN6, RN7} which have widened the scope of future research.\cite{RN8} With the growing interest in soft, flexible, and wearable sensors, a liquid metal-based sensor is proven to be effective in biomedical areas\cite{RN9} including health monitoring, electronic skin, strain sensor, temperature\cite{RN10} and humidity sensor,\cite{RN10} and also an electrochemical sensor for detection of heavy metals ions,\cite{RN11} gases.\cite{RN12}
Among various electrochemical sensing, pH sensing is particularly essential as pH level influences most of the reactions in analytical chemistry, biology, and environmental science. The metal-metal oxide system is a well-known material for solid-state pH sensors.\cite{RN13} The ion-sensitive FET (ISFET) was first introduced by Bergveld\cite{RN15} in 1970 and it has since become mainstream sensor technology for measuring  pH and many different ions, DNA, and biomolecules. The ISFET was further evolved to a \cite{RN16} as extended gate FET(EGFET), which reduces the effort of complex fabrication of ISFET.   E. Mitraka et al. \cite{RN14} investigated a screen-printed hybrid electrode of GaInSn embedded in a conducting polymer (PEDOT) matrix for pH threshold indicator. Naturally, the idea comes to explore the potential of liquid metal featuring superficial oxide as a pH sensor.  To test the merit of liquid metal $Ga-Ga_{2}O_{3}$ system as a pH sensor, we investigated the pH sensing properties, especially the sensitivity and linearity of liquid metal in the form of a pendant drop (LMPD) using MOSFET (metal-oxide-semiconductor field-effect transistor) by performing a series of electrical measurements and analysis in the pH buffer range of 4$\sim$10.
\section*{Electrical Measurement of LMPD}
We measured open circuit potential (OCP) between the LMPD probe referenced to an Ag-AgCl electrode by a precision digital voltmeter (Hewlett Packard HP E2378A) as shown in Figure 1(c). We built an electrical measurement system in a NI-PXIe-1073 chassis with an integrated controller. We installed NI PXI-4132 programmable high-precision source measurement units (SMUs) and a NI-PXIe 6361 DAQ in the NI-PXIe-1073 chassis. A Keithly 2400 low voltage source meter was used as another SMU as two SMUs are required to characterize the three-terminal device i.e., MOSFET(EGFET). We developed a LabVIEW programming interface to control both the SMUs and DAQ. Complete measurement setup, materials, components, and measurement instruments are presented in Figure S1-S2(\textbf{Supporting Information}). The schematic diagram for the I-V measurement of LMPD EGFET is illustrated in Figure S3(\textbf{Supporting Information}). We performed all the electrical characterization of both E-GaInSn and Ga LMPD pH sensors at room temperature (298K) and ambient conditions.
\subsection*{Open Circuit Potential (OCP) Measurement}
Without any transducer (MOSFET), we can easily estimate the intrinsic surface potential in a metal-metal oxide sensor electrode by measuring open circuit potential(OCP)\cite{RN17}. The OCP reflects the true potential of a spontaneous electrochemical reaction at or close to thermodynamics equilibrium. We tested nine (N=9) fresh liquid metal GaInSn drops for measurement in the pH buffer from 4 to 10 and recorded the corresponding stable voltage reading from a precision voltmeter (HP E2378A). We plotted the measured OCPs vs. pH variations with an error bar to estimate the average sensitivity and linearity of the liquid metal sensor probe (Figure 1d). The average sensitivity and linearity are (94.17±13.54) mV $pH^{-1}$ and $R^2$=0.9922±0.0047, N=9, respectively.
\begin{figure*}[t]
    \centering
\includegraphics[width=1\textwidth]{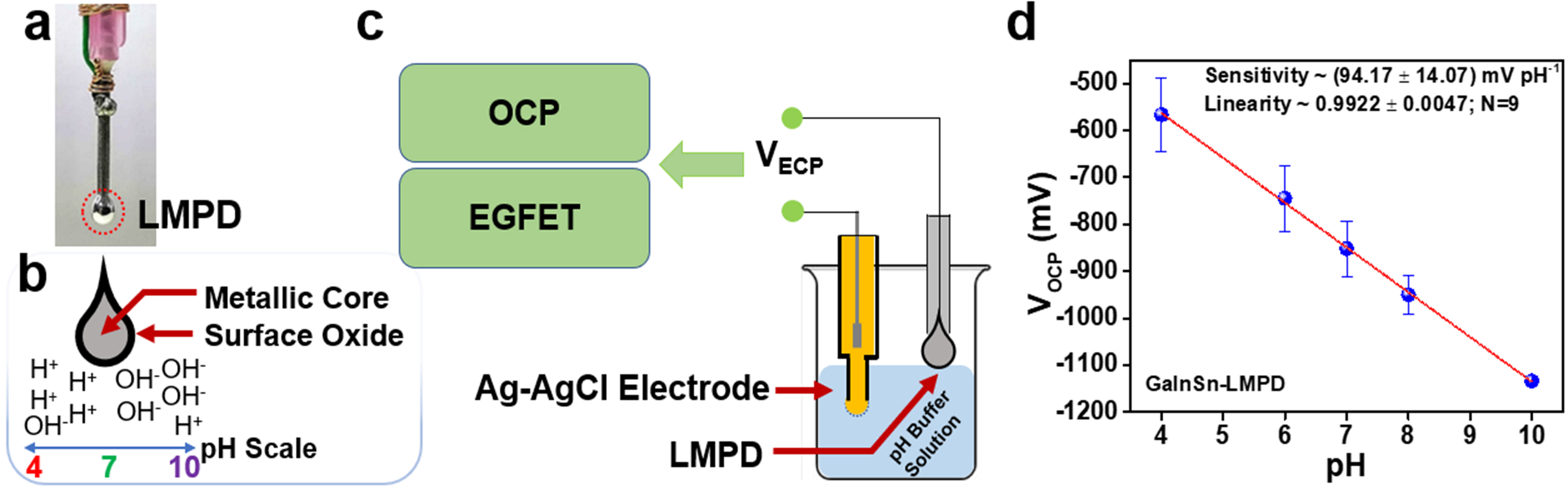}
    \caption{\textbf{LMPD’s pH sensing characterization} a) Photograph of liquid metal pendant drop (LMPD). The liquid metal drop is hanging from the narrow aperture of the stainless needle. b) Schematic diagram to visualize the liquid metal drop in the pH buffer environment. The oxide layer is immune in pH ranges from 4 to 10. Liquid metal comprised of an inner liquid metal core and an ultra-thin natural surface oxide $Ga-Ga_{2}O_{3}$ (0.5$\sim$3 nm). c) Schematic representation of potentiometric measurement setup of LMPD vs. Ag-AgCl electrode using (i) open circuit potential (OCP)  (ii) extended gate (EG)FET d) Measurement result from OCP of nine(9) LMPDs in different pH buffer solutions. The pH sensitivity and linearity were extracted through a linear fit of measured data points.}
    \label{Figure:1}
\end{figure*}
\subsection*{Extended Gate FET (EGFET) LMPD }
The basic MOSFET theory is fully applicable to ISFET or EGFET with a slight modification of threshold voltage term by including the electrochemical cell potential ($E_{ECP}$), which is induced as a surface potential in the sensor probe. The $E_{ECP}$ is the electrode potential which is described satisfactorily by the Nernst equation:
\begin{equation}
E_{\text{ECP}} = E^{0} + \frac{RT}{n_{e^{-}}F} \log(a_{H^{+}}) = E^{0} - \frac{RT}{n_{e^{-}}F} \text{pH} = E^{0} - \frac{0.0591}{n_{e^{-}}} \text{pH}
\end{equation}
$E^0$ is the standard cell potential, R is the universal gas constant, T is the absolute temperature in Kelvin, F is the Faraday constant, $n_{e^-}$ is the number of electron transfers in the process, and $a_{H^+}$ is the activity of hydrogen ion in the solution. The EGFET operated in the linear region $(V_{DS}<V_{GS}-V_{th (MOSFET)})$ the drain current ($I_{DS}$) is expressed as:
\begin{equation}
I_{DS} = \mu_{n}C_{ox}\frac{W}{L}V_{DS}\left[V_{REF} - (V_{th(\text{MOSFET})} \pm E_{ECP}) - \frac{1}{2}V_{DS}\right]
\end{equation}
where $\mu_{n}C_{ox}(W/L)$ is the intrinsic parameter of MOSFET, $V_{REF}$ is the reference bias voltage, $V_{th(MOSFET)}$ is the threshold voltage of MOSFET without any influence of $E_{ECP}$ and $V_{DS}$ is the drain-source voltage. We measured the transfer characteristics of EGFET (GaInSn LMPD) in the linear region for the various pH buffer from 4 to 10 and plotted the curves, as shown in Figure 2a. The curve shifts linearly with increasing (decreasing) pH ($H^+$ ion) values of the buffer solution (pH 4 to 10). We tested fifteen (N=15) fresh liquid metal E-GaInSn drops for EGFET measurements. We extracted the $V_{REF}$ data at 100 µA from each LMPD’s transfer characteristic. We plotted $V_{REF}$ vs. pH values with an error bar (Figure 2b) to estimate the average sensitivity and linearity. The average sensitivity and linearity are (92.96±13.54) mV $pH^{-1}$ and $R^2$=0.9898±0.0122, N=15 respectively. We also tested pure gallium liquid metal drop for further confirmation. We measured the transfer characteristics and estimated the sensitivity of liquid gallium LMPDs. The sensitivity and linearity for a typical gallium LMPD are 110.78 mV $pH^{-1}$ and $R^2$=0.9953, respectively (Figure 2c and Figure 2d).
\begin{figure*}[t]
    \centering
    \includegraphics[width=1\textwidth]{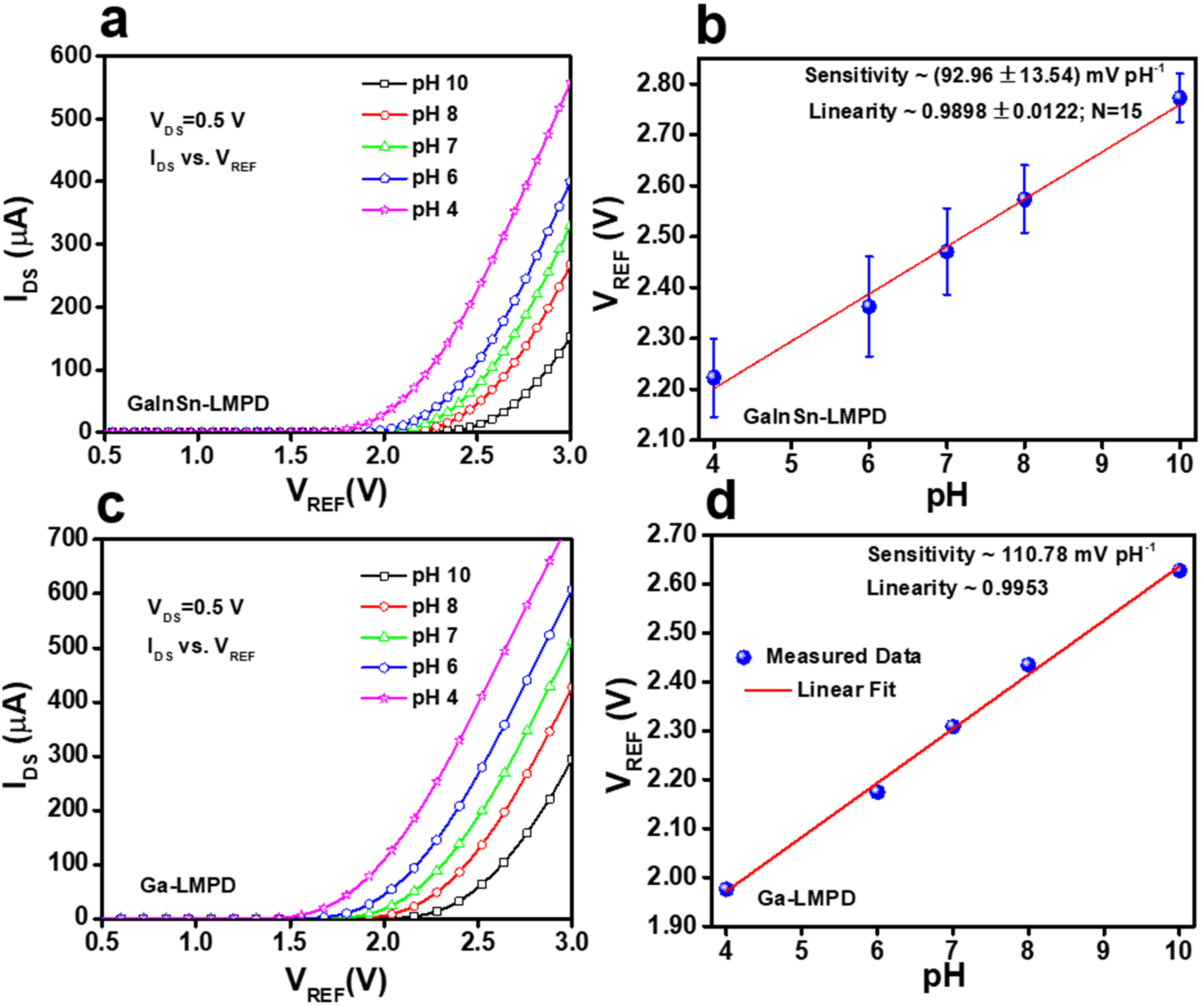}
    \caption{\textbf{LMPD's Electrical characterizations} a) Transfer characteristics of EGFET GaInSn LMPD in different pH buffer solutions from 10 to 4. The LMPD probe responds well to pH change as the curve is gradually shifting towards the right with increasing pH value (decreasing $H^+$ ion concentration). b) Fifteen GaInSn LMPD probes’ $V_{REF}$ at $I_{DS}$=100µA with corresponding pH values plotted with an error bar for estimation of average sensitivity and linearity. c) Transfer characteristics of EGFET Ga LMPD in different pH buffer solutions from 10 to 4. d) For a typical Ga LMPD, $V_{REF}$ vs. pH plot for the sensitivity and linearity calculation.}
    \label{Figure:2}
\end{figure*}
\section*{Discussion}
\subsection*{Origin of Super Nernstian Sensitivity in LMPD}
The sensitivity of a pH sensor is usually related to Nernst's potential of 59.1 mV $pH^{-1}$. The sensitivity of more than 59.1 mV $pH^{-1}$ measured from the pH sensor is often explained with ambiguity. The present understanding of the underlying mechanism of chemical sensing in the solid-state sensor has been carried forward from its inception in 1970\cite{RN15} until now.\cite{RN18} The existing site binding model\cite{RN19} is basically based on Boltzmann distribution which only could explain the classical Nernstian sensitivity of 59.1 mV $pH^{-1}$. The super-Nernstian sensitivity (more than 59.1 mV $pH^{-1}$) obtained from many experiments has been treated as an exception of the classical Nernst equation involving more than one hydrogen ion per one-electron transfer in a solid-liquid interface. The LMPDs give pH sensitivity as low as 71 mV $pH^{-1}$ to as high as 110 mV $pH^{-1}$. It seems a different mechanism plays a role in the solid-liquid interface which could explain both Nernstian and super-Nernstian sensitivity in a unified way. 
\subsection*{Electrochemical Cell and Metal-Metal Oxide Potentiometric Sensor}
The working principle of potentiometric sensors is fundamentally inherited from an electrochemical cell. In a potentiometric sensor, two half-cell reactions take place at each electrode. Only one of the reactions should involve sensing the species of interest and the other should be a well-understood reversible and non-interfering reaction occurring in the Ag-AgCl reference electrode. Usually, metal-metal oxide electrodes respond to pH buffer solution through a reversible electrochemical reaction.\cite{RN13} A generalized redox reaction occurring at the electrode-electrolyte interface can be expressed as follows,

\begin{equation}
M_{x}O_{y}+2yH^{+}+2ye^{-} = xM+yH_{2}O
\end{equation}
The electrode potential can be calculated for the above reaction as follows,
\begin{equation}
E=E^{0}-\frac{\mathrm{0.0591}}{\mathrm{2y}}log[a_{H^+}]^{2y}=E^{0}-(\frac{\mathrm{2y}}{\mathrm{2y}})\times 0.0591log[a_{H^+}]
\end{equation}
Therefore, the sensitivity will be 59.1 mV$pH^{-1}$ at 298K. The possible redox reactions for the $Ga-Ga_2O_3$ system can be written as follows,
\begin{equation}
Ga_2O_3+6H^{+}+6e^{-} = 2Ga+3H_{2}O
\end{equation}
An intermediate oxide $Ga_2O$ can be formed during the reaction, but it is less stable and readily transferred to metallic gallium. These reactions are as follows,
\begin{equation}
Ga_2O_3+4H^{+}+4e^{-} = Ga_2O+2H_{2}O
\end{equation}
\begin{equation}
Ga_2O+2H^{+}+2e^{-} = 2Ga+H_{2}O
\end{equation}
As an equal number of ions ($n_{H^+}/n_{e^-}$) exchange occurs during the process, the sensitivity will be 59.1 mV $pH^{-1}$. Different equilibrium reactions other than those mentioned above (Eq. 5,6,7) only could initiate a sensitivity of more than 59.1 mV $pH^{-1}$. We could write a general electrode reaction with asymmetric ion exchange in the interface involving specific reactants and products as follows:
\begin{equation}
mA+n_{H^{+}}H^{+}+ne^{-} = nB+cH_{2}O
\end{equation}
in which A represents a simple metallic ion or metallic oxide and B is the corresponding metal.
The electrode potential for the above reaction is given below:
\begin{equation}
\begin{array}{r c l}
E&=&E^{0}-\frac{\mathrm{0.0591}}{\mathrm{n_{e^-}}}log\frac{\mathrm{(a_B)^n(a_{H_2O})^c}}{\mathrm{(a_A)^m(a_{H^+})^{n_{H^+}}}}\\&=&E^{0}+0.0591\Bigg(\frac{\mathrm{n_H{^+}}}{\mathrm{n_{e^-}}}\Bigg) log(a_{H^+})-\frac{\mathrm{0.0591}}{\mathrm{n_{e^-}}}log\frac{\mathrm{(a_B)^n(a_{H_2O})^c}}{\mathrm{(a_A)^m}}
\end{array}
\end{equation}
Assuming the activity of A, B, and $H_2O$ to unity, the equation could be simply expressed as follows,
\begin{equation}
E=E^{0}+0.0591\Bigg(\frac{\mathrm{n_{H^+}}}{\mathrm{n_{e^-}}}\Bigg)log(a_{H^+})=E^{0}-0.0591\times x \times pH
\end{equation}
The ion exchange factor ‘‘$x=n_{H^+}/n_{e^-}$’’ will determine the exact electrochemical potential, which will be induced as surface potential in the sensor probe during a spontaneous electrochemical reaction with the analyte. The average sensitivity is 92.96 mV $pH^{-1}$ for E-GaInSn-LMPD and 110.78 mV $pH^{-1}$ for Gallium-LMPD. The value of $x$ is close to 1.57 for E-GaInSn and 1.87 for Gallium.
\subsection*{Pourbaix’s pH-Potential formulation and relation to the $Ga-Ga_2O_3$ System}
Based on the fundamental foundation of thermodynamics and electrochemistry, M. Pourbaix formulated \cite{RN20} the famous pH-potential diagram which could predict for a given element, the equilibrium states of all the possible reactions between the elements, its ions, and its solid and gaseous compound in the presence of water. We found the following reactions of Gallium with asymmetric ion exchange as stated below,
\begin{equation}
Ga+2H_{2}O\leftrightarrow GaO_{2}^-+4H^{+}+3e^{-}
\end{equation}
\begin{equation}
Ga+3H_{2}O\leftrightarrow HGaO_{3}^{2-}+5H^{+}+3e^{-}
\end{equation}
\begin{equation}
Ga+3H_{2}O\leftrightarrow GaO_{3}^{3-}+6H^{+}+3e^{-}
\end{equation}
For these reactions, electrode potential can be written as follows,
\begin{equation}
E=E^{0}-0.0591\Bigg(\frac{\mathrm{4}}{\mathrm{3}}\Bigg) pH +0.0197log(GaO_{2}^-)
\end{equation}
Similarly, the electrode potential for the reaction stated in equations 12 and 13 is as follows,
\begin{equation}
E=E^{0}-0.0591\Bigg(\frac{\mathrm{5}}{\mathrm{3}}\Bigg) pH +0.0197log(HGaO_{3}^{2-})
\end{equation}
\begin{equation}
E=E^{0}-0.0591\Bigg(\frac{\mathrm{6}}{\mathrm{3}}\Bigg)pH + 0.0197log(GaO_{3}^{3-})
\end{equation}
The pH sensitivity calculated from these three equations is 78.8 mV $pH^{-1}$, 98.5 mV $pH^{-1}$, and 118.2 mV $pH^{-1}$, respectively. These values are close to our measured pH sensitivity. The average sensitivity of E-GaInSn and gallium are 92.96 mV $pH^{-1}$ and 110.78 mV $pH^{-1}$, respectively. The values are approaching the theoretical maximum values of 98.6 mV $pH^{-1}$ and 118.2 mV $pH^{-1}$ for the $Ga-Ga_2O_3$ system.  The gallium-based anions ($GaO_3^-$, $GaO_3^{3-}$ and $HGaO_3^{2-}$) are simply behaving as counter-ion and will not contribute to the overall pH response. There are many possible reactions for a given material and its oxides in an aqueous solution. It can be easily understood from the basic principle of thermodynamics and Gibbs's free energy concept. Depending on the materials, the interfacial reaction automatically selects the appropriate reaction to minimize system energy. For example, in our Ga-$Ga_2O_3$ system, the common reactions with symmetric ion exchange  $(n_{H+}/n_{e-}$=6/6=1) or $(n_{H+}/n_{e-}$=3/3=1) will result in a sensitivity of 59.1 mV $pH^{-1}$. The change in the oxidation number of materials involves in the redox reaction controls the number of electron transfers in the interface. As the system will always render to minimize the energy, the next possible reaction occurs at \emph{x}=4/3=1.33, resulting in a sensitivity of 78.8 mV $pH^{-1}$. Similarly, we may get a sensitivity of 98.5 mV $pH^{-1}$ and 118.2 mV $pH^{-1}$ for higher order ion exchange (\emph{x}=5/3, 6/3) at the interface depending on the sensing material's purity. The purity of E-GaInSn and Ga used for the experiment is 99.99 (4N) as shown in Figure S4, and 99.999 (5N) (\textbf{Supporting Information}), and we noticed at least 10 mV $pH^{-1}$ sensitivity differences between two LMPD probes. Here we would like to understand the nature of ‘\emph{x}’. The $n_{H^+}$ is the number of hydrogen ions adsorbed in the sensor surface, and its values are 1, 2, 3, … n. The $n_{e^-}$ is the change of oxidation number of the material involved in the reactions. Therefore, ‘\emph{x}’ has a precise step-wise discrete value, which in turn control sensitivity in a similar manner.  Even if the sensitivity is approaching the theoretically predicted value, differences arise between the theoretical and experimental sensitivities owing to the deviation of material electronic configuration from its ideal conditions. The ideal electronic configuration of the material may perturb due to strain, impurity, defects, scattering, etc. Here we introduce an entirely new parameter, Pourbaix factor ‘pb’ (in honor of  M. Pourbaix’s work), which quantitatively represents the deviation of materials quality from its ideal condition. 
The new Nernst equation by including ‘pb’ can be written as follows,
\begin{equation}
E=E^{0}-0.0591\times pb \times x \times pH
\end{equation}
The value of ‘pb’ will be between 0 and 1 ($0<pb<1$).
\subsection*{Validation of Pourbaix factor (pb) in Gallium Nitride (GaN)-$Ga_2O_3$ System}
In the case of a gallium nitride system, the native oxide is also $Ga_2O_3$. Why does such a system never exhibit a sensitivity of more than 59.1 mV $pH^{-1}$? The system is already perturbed by the incorporation of nitrogen (N) into Gallium (Ga), which leads to imperfection in native $Ga-Ga_2O_3$. We estimated the ‘pb’ for the $GaN-Ga-Ga_2O_3$ system based on the lattice parameters of Ga\cite{RN21} and GaN.\cite{RN22} The pb can be calculated from the expression stated below,
\begin{equation}
pb=1-\Bigg[\frac{\mathrm{a_{Ga}-a_{GaN}}}{\mathrm{a_{Ga}}}\Bigg]
\end{equation}
 As the system will always try to minimize its energy, the next possible reaction after \emph{x}=3/3 (59.1) for $Ga_2O_3$ will be for \emph{x}=4/3( 78.8 mV $pH^{-1}$). Due to imperfection in $Ga_2O_3$, the actual sensitivity will be 78.8×0.7046=55.60 mV $pH^{-1}$ (considering lattice constant ‘a’) and 53.41 mV $pH^{-1}$ (considering lattice constant ‘c’).  M. Bayer et al.\cite{RN23} calculated  a sensitivity of  55.90 mV $pH^{-1}$ using density functional theory (DFT) which is similar to the sensitivity we derived considering the ‘pb’ illustrated in Table~\ref{tbl: pb}.
Based on experimental and theoretically predicted pH sensitivities, we introduce a compact plot between  sensitivity (S) vs. Pourbaix-ion exchange (pbx) factor to represent all the possible sensitivity values for the Ga-$Ga_2O_3$ system (Figure 3). The family of lines with decreasing slopes represents the sensitivity with different Pourbaix factors (pb).
\begin{table*}
  \caption{Pourbaix factor in $GaN-Ga_2O_3$ system: Parameters for calculation of ‘pb’ in $GaN-Ga_2O_3$ system and associated pH sensitivity}
  \label{tbl: pb}
  \centering
  \begin{tabular}{c c c c c c c c}
    \hline
   & a(\AA) & c(\AA) & $pb_a$ & Sens. & $pb_c$  & Sens.& Sens.(Theo.)\cite{RN23}  \\
    \hline
    Ga   & 4.5258 & 7.6570   \\
         &        &       & 0.70 &  -55.60 & 0.67 & -53.41 & -55.90   \\
    GaN & 3.1890 & 5.1864 \\
    \hline
  \end{tabular}
 \end{table*}
\begin{figure*}[t]
    \centering
    \includegraphics[width=0.9\textwidth]{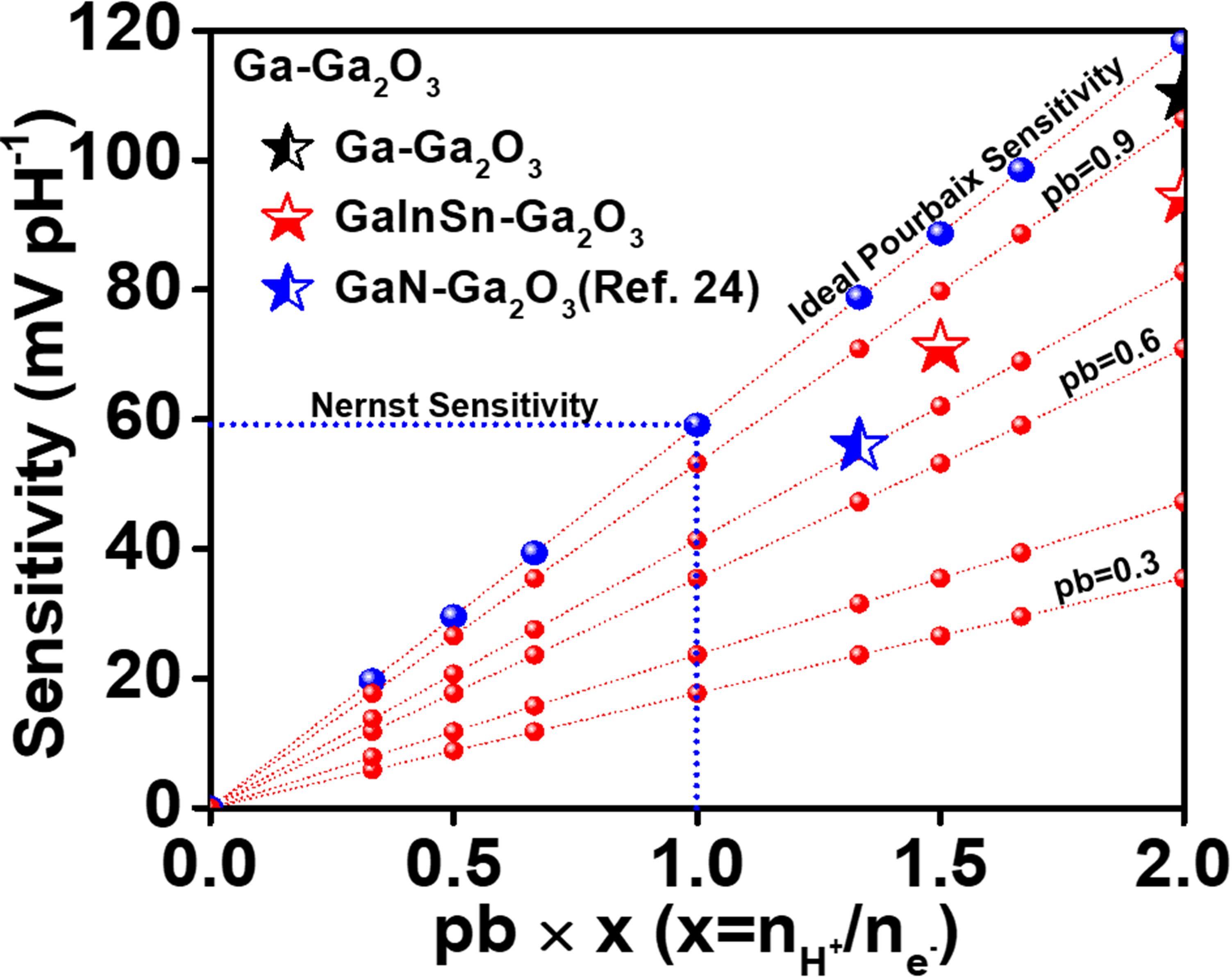}
    \caption{\textbf{Sensitivity vs. compact Pourbaix-ion exchange factor (S-pbx) plot for $Ga-Ga_2O_3$ material systems}: Sensitivity vs. ion exchange factor (Pourbaix factor) plot to visualize all possible sensitivity values which could be obtained through different pbx factors. Based on our experimentally obtained highest sensitivity in the $Ga-Ga_2O_3$ system, the pbx is set to 2 to accommodate all the sensitivity values. Available experimental sensitivities from the literature are plotted in terms of the Pourbaix factor (pb), which give a sense of deviation of the material quality from its ideal one.}
    \label{Figure:3}
\end{figure*}
\subsection*{Critical Discussion}
Nernstian sensitivity is estimated by assuming a one-electron exchange through the interface. However, it is the ion exchange factor ‘\emph{x}’=1, which determines the sensitivity of 59.1 mV $pH^{-1}$. Depending on the value of ‘\emph{x}’, the sensitivity will evolve from sub-Nernstian $(S<59.1,   \emph{x}<1)$ to Nernstian ($S=59.1, \emph{x}=1$) to super-Nernstian ($S> 59.1, \emph{x}>1$) regime. We propose it as ideal Pourbaix sensitivity as the occurrence of lower and higher-order (except \emph{x}=1) ion exchange in the interface was originally proposed by M. Pourbaix. Why such high sensitivity is not usually accessible for thick metal-metal oxide or metal-oxide sensing film?  For thick oxide, the higher-order ion exchange is not likely possible as the sensing film’s resultant electronic configuration largely deviates from its ideal monolayer’s electronic configuration. Natural oxide with a high degree of perfection only can exhibit a higher order of ion exchange in the interface during electrochemical reactions, which is mostly missing in the case of thick oxide grown on the arbitrary substrate. In quasi-2D or 2D oxide, the resultant electronic configurations, energy levels, chemical bonding, and hence the interfacial interactions with the analyte are approaching the ideal theoretical limit of the elemental oxide itself. The limitation caused by the structural features of bulk material are alleviated in 2D film and the intrinsic material properties are overwhelmingly effective by the resulting properties in the 2D structure. In E-GaInSn (gallium) energetically favorable self-limiting, amorphous (poorly crystalline) $Ga_2O_3$ is formed in the metal-air interface, which is atomically thin and extraordinarily smooth naturally grown 2D oxide.\cite{RN2,RN3} Moreover, the $Ga_2O_3$ is relieved from the strong covalent bonding in an orthorhombic crystal structure upon its formation from liquid Gallium, and a weak van der Waals (vdW) force only exists between oxide and metallic core. Therefore, the higher-order ion exchange is very susceptible to the E-GaInSn-$Ga_2O_3$ system, which leads to high sensitivities.
\subsection*{Conclusion}
Ever-increasing research in recent times on liquid metal in the field of flexible and wearable sensors motivated us to study the pH-sensing properties of liquid E-GaInSn-$Ga_2O_3$ systems. The high sensitivity in the liquid metal system originated from the higher-order ion exchange $(\emph{x}>1)$ occurring in the ultra-thin oxide-pH interface. The whole ion sensing process is a spontaneous interfacial electrochemical reaction well described by Gibbs, Nernst, and Pourbaix equations. We found that Nernstian sensitivity of 59.1 mV $pH^{-1}$ occurs only for symmetric ion exchange (\emph{x}=1) reactions. Asymmetric ion exchanges ($\emph{x}\neq1$) could result in sensitivity far beyond the Nernst sensitivity, depending on the properties of the material. We derived and validated a generalized equation by including ion exchange ‘\emph{x}’ and the Pourbaix factor (pb) in the well-known Nernst equation, which could explain almost all sensitivity values available in the literature. High sensitivity ($S>59.1 mVpH^{-1}$) is usually inaccessible in the thicker oxide owing to the intrinsic material and electronic properties.  However, in a quasi-2D or 2D oxide, the electronic interaction is approaching the elemental monolayer oxide level resulting in an enhanced interaction in the interface; hence the high sensitivity is easily attainable. Our study on the liquid metal system paves the way for developing next-generation electrochemical sensors based on 2D oxide.
\section*{Experimental}
\subsection*{Materials}
Eutectic gallium alloy E-GaInSn (Galinstan) and gallium (Ga) with purity 99.99(4N) and 99.999 (5N) were purchased from RICH-METALS Inc. China.  The NIST standard pH buffer solutions of pH 4, 6, 7, 8, and 10 were purchased from Alfa Aesar Inc. and Hanna Instruments Inc. A double junction Ag-AgCl (OHAUS STREF1) reference electrode was purchased from Analytical Instruments, Inc, China. An n-type enhancement mode MOSFET in a commercial IC CD4007UBE was used as the electronic transducer for EGFET.
\subsection*{Method}
We filled liquid E-GaInSn in a 1 mL syringe fitted with a stainless-steel dispensing needle with a diameter of 0.6 mm. The liquid metal in the syringe would come out from the pinhead aperture of the needle with a gentle push of the syringe’s piston. Careful control of the piston movement would form a drop of liquid metal suspended from the pinhead aperture, which we call ‘Liquid Metal Pendant Drop’ (LMPD) as shown in Figure 1(a).  The hanging electrode concept was adapted from a polarography experiment date back in 1954 by P. A. Giguère et al.\cite{RN24} We also prepared gallium LMPD for the pH response measurement.  Initially, gallium was melted in a temperature bath maintained at 35\textdegree C, and we filled some molten liquid gallium in a syringe (same preparation as E-GaInSn LMPD). Because of the super-cooling effect, the molten gallium maintains its liquid state even if the temperature is lower than 29.76\textdegree C (melting temperature of Ga). We soldered a wire on the stainless needle towards the syringe side, which provided the electrical contact between the LMPD probe and the MOSFET gate.



\section*{Acknowledgments}
We would like to thank Professor Lee Chow, University of Central Florida, for his valuable scientific input and fruitful discussion regarding the work. The author (A.D) would like to thank Dr. Yu Cheng Chang, Aalto University, for the LabVIEW programming support in the initial stage of the work. ORCID Atanu Das 0000-0002-1733-626X. Data Statement: All data are reported in the paper or supplementary materials. Dedication: A.D. would like to dedicate this research article to his parents (Kananbala Das and Nagendranath Das), who have been a constant source of inspiration and have given him the drive and discipline to tackle any task with enthusiasm and determination.
\bibliography{Manuscript}
\bibliographystyle{unsrt}
\section*{Supporting Informations}
\setcounter{figure}{0}
\renewcommand{\thefigure}{S\arabic{figure}}
The supporting information is provided to strengthen the experimental section of the main manuscript and it is available free of charge.
\begin{itemize}
  \item Figures S1--S3
  \item Certificate of Analysis (COA) of GaInSn(4N) and Ga(5N)
   \item Movie 1: Liquid Metal Pendant Drop (LMPD) gently manipulated using a height meter with micrometer precision for immersion into the various pH buffer during measurement.
   \item Movie 2: OCP measurement using an HP-E2378A multimeter of EGaInSn LMPD in pH 10 buffer solution.
   \item Movie 3: Transfer characteristics ($I_{DS}-V_{REF}$) measurement of EGaInSn LMPD using Keithly and NI SMUs in pH 10 buffer solution.
  \begin{figure*}[t]
    \centering
    \includegraphics[width=0.95\textwidth]{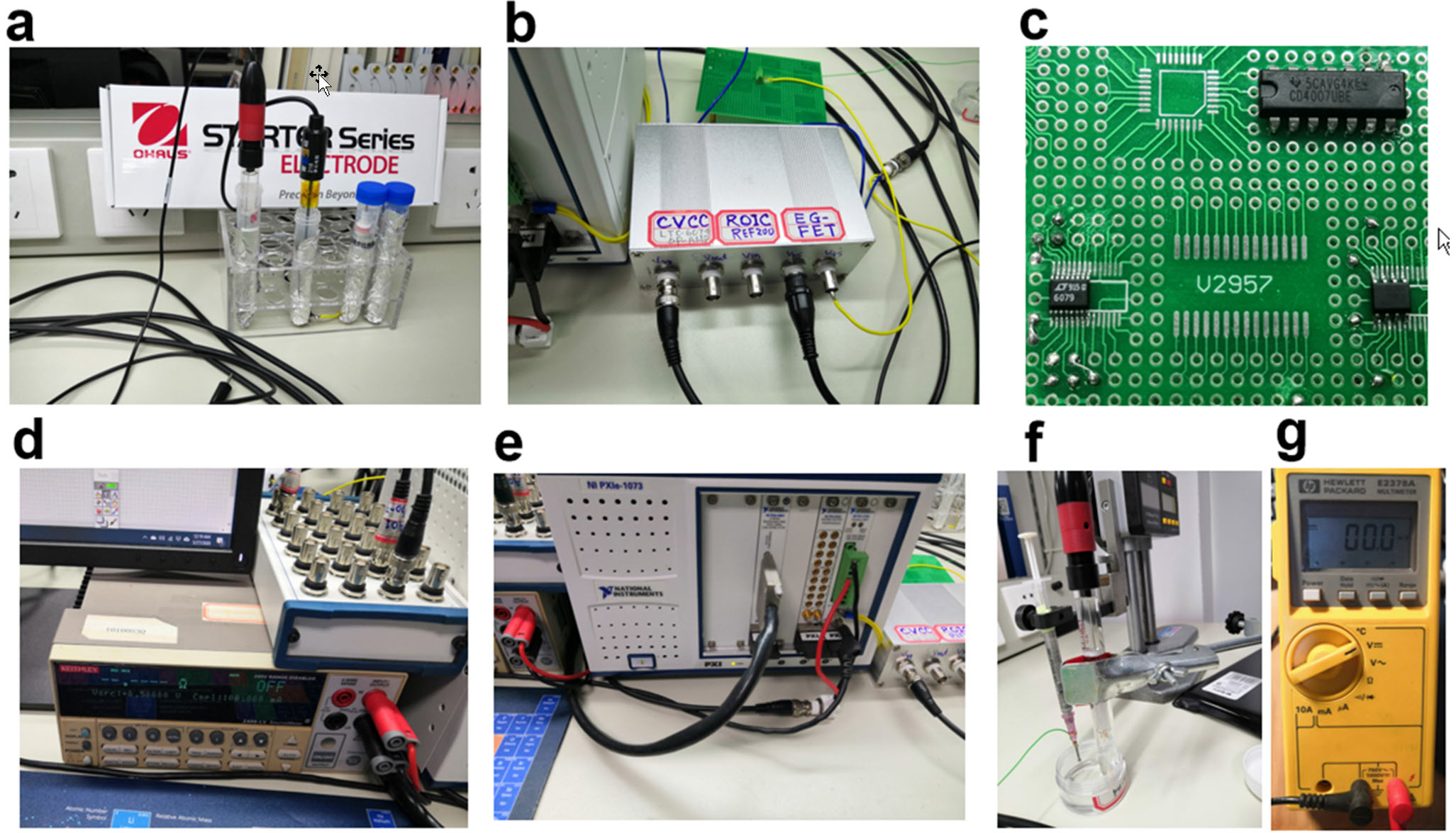}
    \caption{\textbf{The highlight of individual components for experimental procedures}: Photograph of a) the Ag-AgCl reference electrodes. The sensor probe was measured mainly through a double junction reference electrode (OHAUS STREF-1) b) Homemade transducer box equipped with a CVCC-ROIC and EGFET. Input-output leads are connected using BNC connectors for ease of connecting to the measurement instruments’ port c) Photograph of the part of PCB showing the three important components (CD4007UBE on the top, LTC 6079 on the left, and REF200 on the right) d) Keithly 2400 LV source-meter as SMU-2 (mainly use it for measuring $I_{DS}$ and sweeping $V_{DS})$. e) NI PXIe-1073 Chassis equipped with a NI-PXIe-6361 DAQ (mainly use DAQ for CVCC ROIC’s output voltage measurement), NI-PXI 4132 as SMU-2 (mainly use it for $V_{GS}$ and $V_{REF}$ bias sweep.) f) Two electrode setup for potentiometric measurement, (left) LMPD probe and (right) Ag-AgCl electrode. g) Precision HP E2378A multimeter used for OCP measurement (Movie S2).}
    \label{Figure:S1}
\end{figure*}
\begin{figure*}[t]
    \centering
    \includegraphics[width=0.95\textwidth]{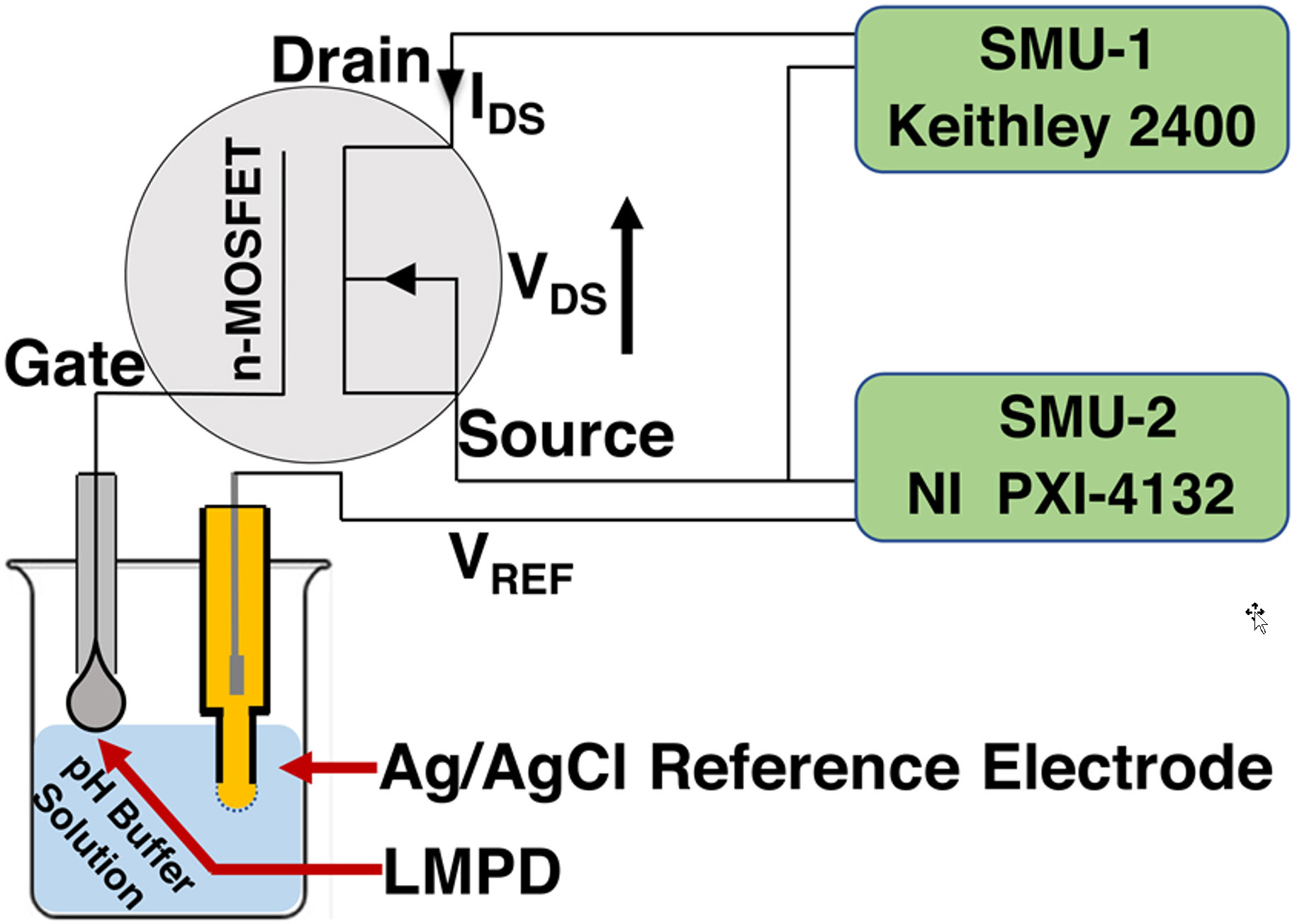}
    \caption{\textbf{Schematic illustration of the LMPD-EGFET measurement setup}: SMU-1 is assigned to keep drain to source voltage $(V_{DS}$) constant at 0.5 V while at the same time measuring the source to drain current $(I_{DS})$. SMU-2 is assigned to sweep the voltage of the Ag-AgCl reference electrode ($V_{REF}$). With the synchronization of both SMUs, we measured the transfer characteristics $(I_{DS}$ vs.$V_{GS})$ of EGFET in various pH buffer solutions (Movie S3). The $E_{ECP}$ induced in the LMPD probe (liquid gate) is reflected in the MOSFET gate to the source terminal; hence the parallel shift occurs in transfer characteristics with different pH.}
    \label{Figure:S2}
\end{figure*}
\begin{figure*}[t]
    \centering
    \includegraphics[width=0.68\textwidth]{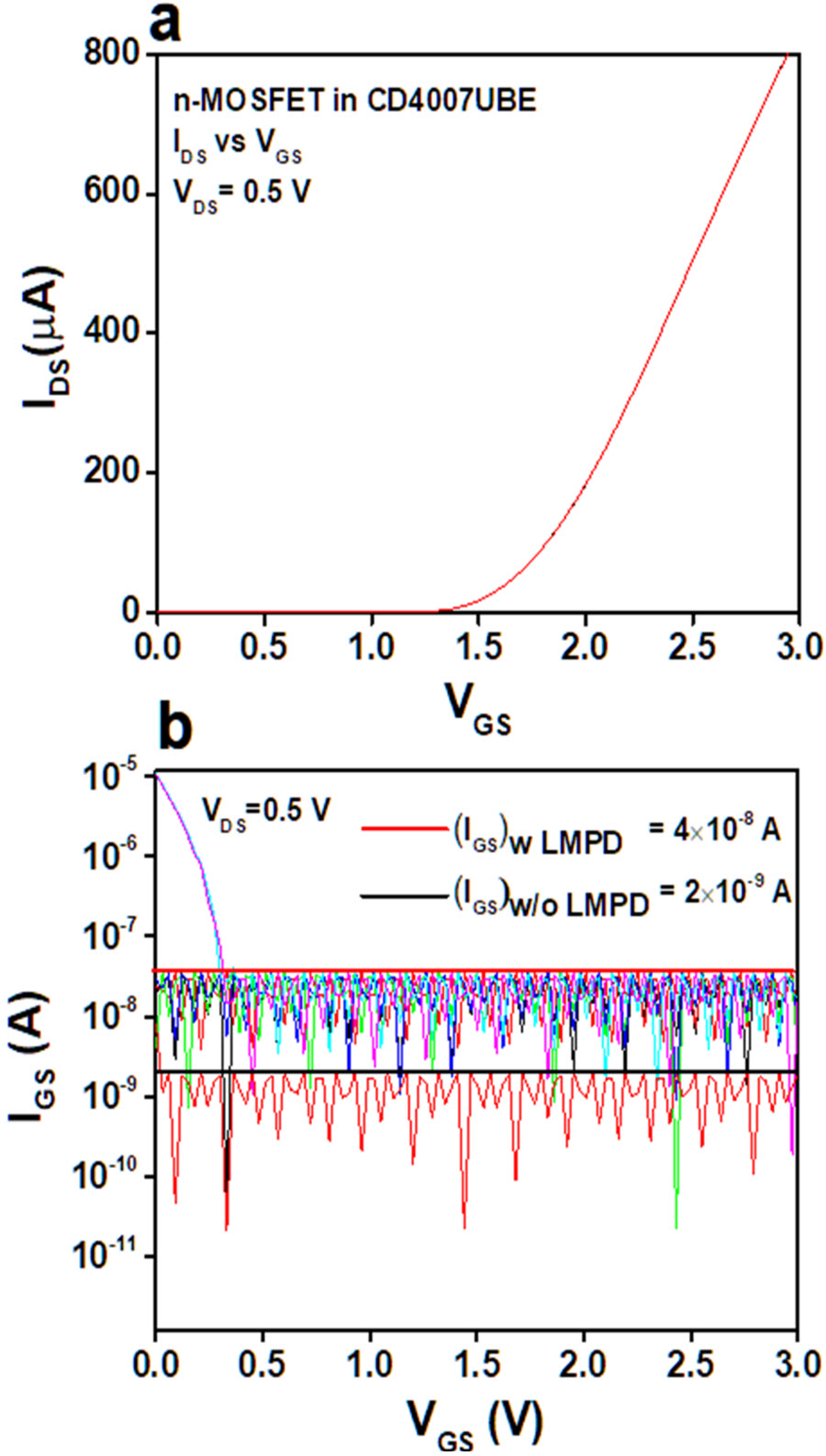}
    \caption{\textbf{Test of transfer and gate leakage characteristics of n-MOSFET}: a) Transfer characteristics ($I_{DS}$ vs. $V_{GS}$) of n-MOSFET in the linear region by suitably controlling $V_{DS}$=0.5 V. The same n-MOSFET is used as an extended gate FET for the LMPD sensor probe. b) Test of gate leakage current ($I_{GS}$) while LMPD is attached to the gate terminal of n-MOSFET. Gate leakage current of n-MOSFET measured for comparison. Gate leakage current is nearly one order higher with LMPD than the n-MOSFET configuration.}
    \label{Figure:S3}
\end{figure*}
\begin{figure*}[t]
    \centering
    \includegraphics[width=0.98\textwidth]{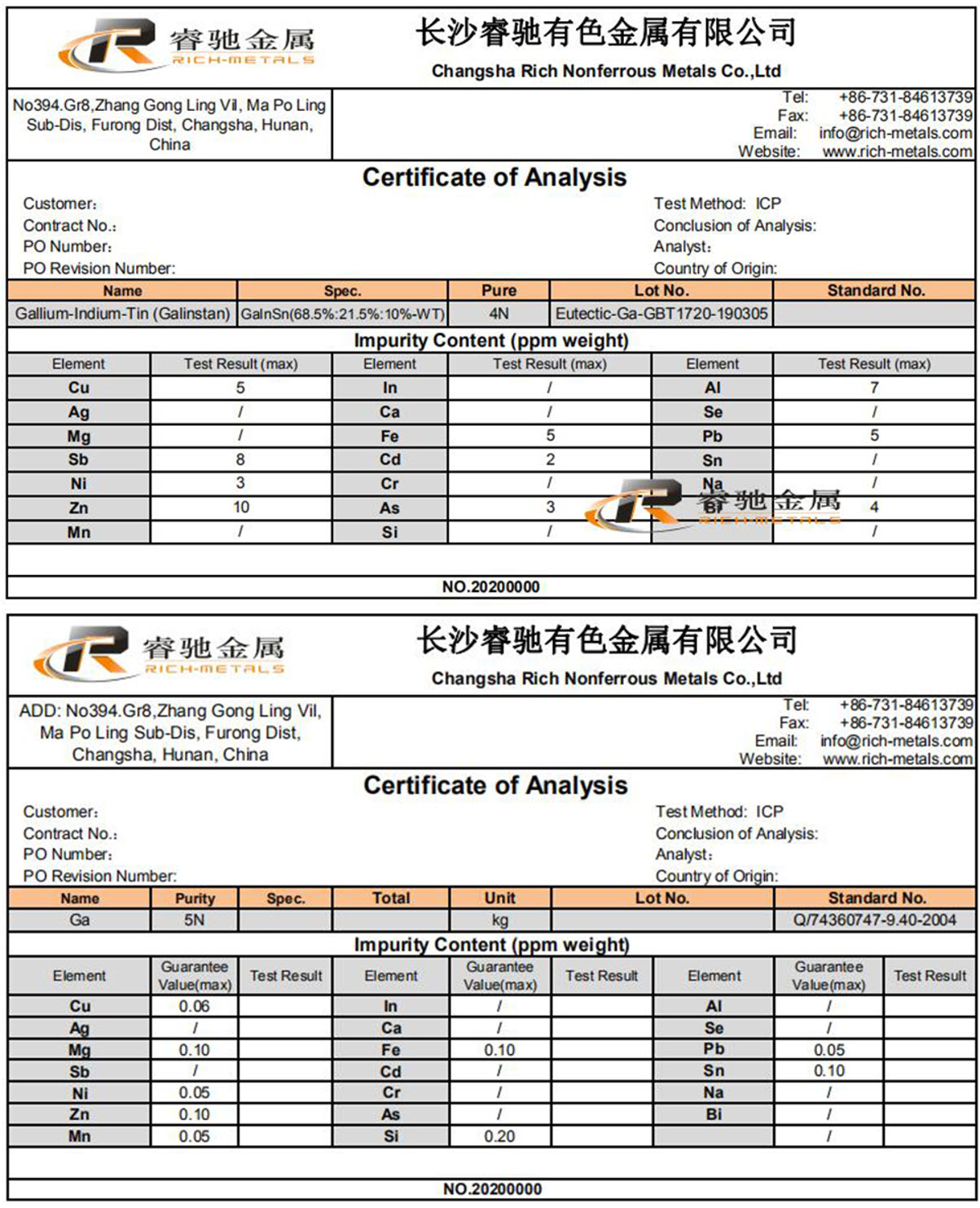}
    \caption{Certificate of Analysis (COA) of GaInSn (4N) and Ga (5N) provided by RICH-Metals, Inc., China}
    \label{Figure:S4}
\end{figure*}
\end{itemize}

\end{document}